\documentclass[AMA,Times1COL]{WileyNJDv5}

\articletype{Special Issue Article}%

\received{Date Month Year}
\revised{Date Month Year}
\accepted{Date Month Year}
\journal{Journal}
\volume{00}
\copyyear{2026}
\startpage{1}
\usepackage{subcaption}
\usepackage{multicol}
\usepackage{comment}

\begin{document}

\title{Short-Horizon Sparse Model Predictive Control for Precipitation Reduction Using Numerical Weather Prediction}

\author[1]{Yuta Tanikawa}

\author[1]{Yuga Tomita}

\author[1]{Toshiyuki Ohtsuka}

\authormark{TANIKAWA \textsc{et al.}}
\titlemark{MODEL PREDICTIVE CONTROL...}

\address[1]{\orgdiv{Graduate School of Informatics}, \orgname{Kyoto University}, \orgaddress{\state{Kyoto}, \country{Japan}}}



\corres{Department of Informatics, Graduate School of Informatics, Kyoto University, Kyoto 606-8501, Japan. Email: ohtsuka@i.kyoto-u.ac.jp}



\fundingInfo{JST Moonshot R\&D Program Grant Number JPMJMS2389-1-1}

\abstract[Abstract]{Extreme weather events exacerbated by climate change demand the development of quantitative weather control technologies for disaster mitigation. 
This study proposes a precipitation control framework integrating a realistic Numerical Weather Prediction (NWP) model with model predictive control (MPC). 
At each control instant in MPC, a finite-difference sensitivity matrix is constructed from the NWP model and used as a local linear model of how perturbations to the atmospheric state affect future precipitation. 
A sparse convex optimization problem is then solved to compute the control input, which is implemented as a perturbation to the atmospheric state. 
To reduce computational cost in sensitivity analysis, multiple grid points in the NWP model are treated collectively as a single block, and a uniform perturbation is applied to all points within each block. 
Moreover, a tailored convex optimization problem is introduced to effectively control the accumulated precipitation at the end of a weather event, using a prediction horizon much shorter than the entire event duration while promoting spatially sparse atmospheric perturbations. 
To evaluate the proposed MPC method, four control methods are compared: (i) initial-only open-loop optimal control (IO-OL), (ii) full-horizon open-loop optimal control (FH-OL), (iii) shrinking-horizon optimal control (SHOC) with a fixed terminal time, and (iv) single-move MPC with a fixed prediction-horizon length. 
Numerical experiments on a warm bubble benchmark demonstrate that MPC achieves precipitation reduction comparable to SHOC while reducing the total computational time relative to FH-OL and SHOC.
Moreover, despite using a linear prediction model, MPC successfully achieves a challenging level of precipitation reduction, even when open-loop optimal control methods, namely, IO-OL and FH-OL, fail because of nonlinear atmospheric evolution. 
These findings suggest that MPC is a promising control framework for NWP-based precipitation reduction in complex weather events. 
} 


\jnlcitation{\cname{%
\author{Tanikawa Y.},
\author{Tomita Y}, and
\author{Ohtsuka T}}.
\ctitle{On simplifying ‘incremental remap’-based transport schemes.} \cjournal{\it J Comput Phys.} \cvol{2021;00(00):1--18}.}

\maketitle

\renewcommand\thefootnote{}

\renewcommand\thefootnote{\fnsymbol{footnote}}
\renewcommand{\vec}[1]{\boldsymbol{#1}}
\setcounter{footnote}{1}

\section{Introduction}\label{sec1}

Extreme windstorms and torrential rainfall events pose increasing risks to human lives, infrastructure, and economic activity. In Japan, particular attention has been paid to typhoons and linear precipitation zones (LPZs), whose spatial localization and rapid development often lead to severe hydrometeorological disasters. In response to these risks, the Japanese Moonshot Research and Development Program Goal 8 promotes research toward weather control technologies that can change the intensity, timing, and location of typhoons and torrential rains for disaster mitigation.\cite{JSTMoonshotGoal8,JBoyd} 
This national research direction has stimulated renewed interest in the mathematical, computational, and physical foundations of weather control.

Weather modification itself has a long history, particularly in the context of cloud seeding for precipitation enhancement.\cite{Flossmann2019BAMS} 
More recently, numerical weather prediction (NWP) models have been used to examine physically motivated intervention mechanisms for mitigating heavy rainfall. For example, Hiraga et al.\cite{Hiraga2026NHESS} investigated idealized cloud-seeding experiments for a mesoscale convective system in Japan and showed that cloud seeding can redistribute rainfall, reducing rainfall in the heavy-rainfall region while increasing rainfall downwind. Such studies provide valuable insight into possible physical intervention mechanisms. However, they do not directly solve the optimal control problem of determining when, where, and how strongly interventions should be applied so that a desired precipitation distribution is achieved while avoiding unintended effects in surrounding regions.

In parallel with these physically motivated studies, mathematical and computational approaches to weather control have been developed.\cite{Sawada} 
A representative framework is the control simulation experiment (CSE), which extends the idea of observing system simulation experiments in meteorology by applying small perturbations to a simulated atmospheric system in order to guide a chaotic system toward a desirable state. Miyoshi and Sun\cite{Miyoshi2022NPG} introduced this concept using the Lorenz-63 system, and Sun et al.\cite{Sun2023NPG} extended it to the mitigation of extreme events in the Lorenz-96 model. Related ideas also appear in variational approaches to severe-weather control and exigent forecasting, where four-dimensional variational data assimilation is used to identify perturbations that influence subsequent weather evolution.\cite{Henderson2005QJRMS} 
These studies demonstrate the conceptual feasibility of controlling chaotic atmospheric dynamics. However, their extension to realistic NWP-based precipitation control remains computationally challenging because of the high dimensionality, strong nonlinearity, and expensive simulation cost of atmospheric models.

Recent optimization-based studies have addressed this difficulty from two complementary directions. One direction treats the NWP model as a black-box simulator and searches for effective interventions without requiring gradient information. 
Higuchi et al.\cite{Higuchi2026JCS} developed and evaluated a black-box optimization framework for weather intervention design, comparing derivative-free optimization methods under limited NWP evaluations. Such black-box approaches are attractive when adjoint or sensitivity information is unavailable. 
Another direction numerically constructs sensitivity matrices from NWP simulations and formulates the intervention design as a convex optimization problem. 
Ohtsuka et al.\cite{ohtsuka} proposed a convex optimization method based on a sensitivity matrix to determine sparse initial atmospheric perturbations for quantitative weather control. 
Kim and Ohtsuka\cite{KIM} extended this idea to perturbation time series, allowing open-loop interventions at multiple control instants. These sensitivity-based convex formulations are well-suited to imposing spatially distributed precipitation constraints and sparsity-promoting penalties, which are important for localized and interpretable interventions.

However, sensitivity-based open-loop strategies have an inherent limitation. Since all perturbations are computed in advance using a linear model around a nominal trajectory, they cannot compensate for nonlinear error growth after interventions are applied. This limitation becomes particularly serious when the desired reduction in precipitation is large or when the control period is long. 
To mitigate this issue, Tanikawa and Ohtsuka\cite{Tanikawa2025JACC} proposed a shrinking-horizon optimal control with a fixed terminal time that sequentially recomputes perturbations using the latest atmospheric state, resulting in closed-loop control. 
Although closed-loop control improves robustness to nonlinearities in state evolution, repeatedly constructing sensitivity matrices over the entire remaining simulation horizon can be computationally expensive for NWP-based weather control.

Model predictive control (MPC) provides a natural framework for resolving the above-mentioned trade-off between closed-loop performance and computational cost. 
In MPC, a finite-horizon optimal control problem is solved at each control instant using the latest state information, and only the first control move is applied to the system; this is a form of closed-loop control.\cite{Mayne2000Automatica} 
If the prediction horizon in MPC is shorter than the entire remaining simulation horizon, MPC requires a lower computational cost than the shrinking-horizon optimal control with a fixed terminal time, possibly at the expense of closed-loop performance. 
Therefore, one of the key issues in MPC for weather control is formulating an optimal control problem that achieves the best possible control performance with the shortest possible prediction horizon. 

MPC has recently been introduced into weather-control-related CSEs and chaotic-system control. Kawasaki and Kotsuki\cite{Kawasaki2024NPG} combined MPC with data assimilation to guide the Lorenz-63 system toward a prescribed regime. 
Sawada\cite{Sawada2} discussed the close relationship between ensemble Kalman filtering and MPC and proposed an ensemble-based control approach for chaotic systems. Kurosawa et al.\cite{Kurosawa2025NPG} further developed ensemble-based MPC using data assimilation techniques for high-dimensional nonlinear systems. These studies indicate that MPC is a promising framework for controlling chaotic geophysical systems under uncertainty and computational constraints. 
However, applications of MPC to NWP-based sparse precipitation control, where sensitivity calculations dominate the computational cost, remain limited. 

Motivated by the above situation, this study proposes an NWP-based sparse MPC framework for quantitative precipitation control. 
In contrast to the shrinking-horizon optimal control with a fixed terminal time, the proposed method evaluates precipitation only over a fixed short prediction horizon. 
At each control instant, a finite-difference sensitivity matrix is constructed over this prediction horizon, and a convex optimization problem is solved to determine the current atmospheric perturbation as the control input, resulting in closed-loop control. 

The main contributions of this study are summarized as follows. 
\begin{enumerate}
 \item We formulate a sparse convex optimization problem for MPC to achieve quantitative precipitation control specifications with a short prediction horizon and a single control move at each control instant. 
 \item We also formulate tailored sparse convex optimization problems for open-loop and closed-loop optimal control to compare with MPC in terms of control performance and computational time. 
 Although they have already been applied to precipitation control, the formulations presented in this paper differ from those in previous studies. 
 \item We compare the proposed MPC with initial-only optimal control, full-horizon open-loop optimal control, and shrinking-horizon optimal control with a fixed terminal time in a warm bubble benchmark. The results show that the proposed MPC achieves precipitation reduction comparable to the shrinking-horizon optimal control while substantially reducing the total computational time. 
\end{enumerate}
These findings suggest that short-horizon sparse MPC is a promising computational approach for online weather control, although further acceleration and validation in realistic three-dimensional cases are required.

The remainder of this paper is organized as follows. 
Section \ref{sec2} summarizes the NWP model and the experimental setup in this study. 
Section \ref{sec3} introduces four control methods for precipitation reduction. 
Section \ref{sec4} presents the experimental results and discusses their implications. 
Finally, Section \ref{sec5} concludes this paper.

\section{NWP Model and Experimental Setup}\label{sec2}

An NWP model predicts future atmospheric conditions through numerical integration of the governing physical equations. 
These equations consist of a system of nonlinear coupled partial differential equations representing the conservation of mass, momentum, and energy (the thermodynamic equation) across various dynamical and physical processes, including cloud microphysics. 
Given an initial atmospheric state, which comprises variables such as temperature, humidity, and wind speed, an NWP model simulates weather phenomena by computing the time evolution of these variables. This allows for the acquisition of meteorological data, including accumulated precipitation, at the end of the simulation.
This study employs SCALE-RM ver.\ 5.5.3 \cite{Nishizawa, Sato} as the NWP model for numerical experiments. SCALE-RM is a regional non-hydrostatic model widely adopted in various meteorological research studies \cite{Honda}. The model provides a variety of physical schemes for each of its components; in this study, a 6-class single-moment bulk scheme \cite{Tomita} is used for cloud microphysics. For simplicity, the effects of turbulence and thermal radiation are omitted from the current model configuration.

In this study, the control input represents the amplitude of the perturbation applied to the atmospheric state in a specific domain at a specific time instant. 
Specifically, the perturbed state variable is the density-weighted potential temperature $\rho\theta$. 
Although intervention technology to perturb the real atmospheric state is still under development and has not been established, it is assumed that the density-weighted potential temperature can be perturbed as the control input for the conceptual study of control methods in this paper. 
Let $\vec{X}_t \in \mathbb{R}^{n_{var}N}$ be the vector of $n_{var}$ atmospheric state variables (density, momentum components, density-weighted potential temperature, and water-substance variables) over $N$ grid points in the domain of interest at a given discrete time step $t \in \{0,1,\ldots,T-1\}$. 
The discrete time step represents the index of control instant for observing and perturbing the atmospheric state. 
Here, the simulation terminal time $T$ is fixed to evaluate the accumulated precipitation at the end of a specific weather event. 
The components of the vector $\vec{X}_t$ are ordered in a suitable way by rearranging the atmospheric state variables in the domain of interest. 
The accumulated precipitation over a time interval from time step $t_1$ to time step $t_2$ $(0 \le t_1 < t_2 \le T)$ is denoted by $\vec{y}_{t_1:t_2} \in \mathbb{R}^m$. 
Then, the input-output map of the NWP model from $t_1$ to $t_2$ can be regarded as a nonlinear function $F_{t_1:t_2}^{sim}$, expressed as follows:
\begin{equation}
    \vec{y}_{t_1:t_2} = F_{t_1:t_2}^{sim}(\vec{X}_{t_1}).
\end{equation}
The function $F_{t_1:t_2}^{sim}$ is determined by the NWP model, which numerically solves the governing physical equations from $t_1$ with the initial condition $\vec{X_{t_1}}$ up to $t_2$. 

\begin{figure}[t]
    \centering
    \includegraphics[width=0.5\linewidth]{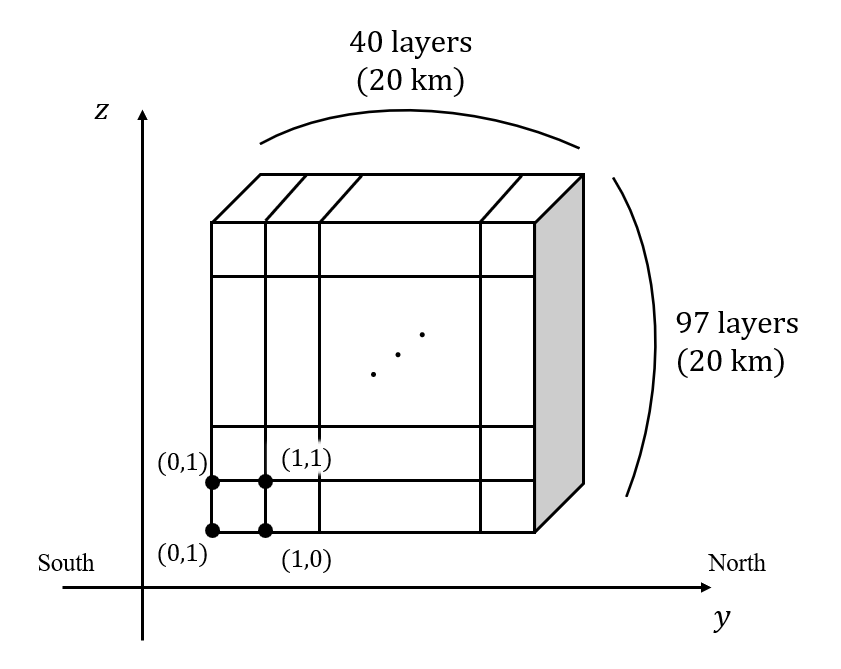}
    \caption{Grid configuration and environmental conditions for the warm bubble experiment.}
    \label{fig:grid}
\end{figure}

\begin{figure*}[t]
    \centering
    \includegraphics[width=\linewidth]{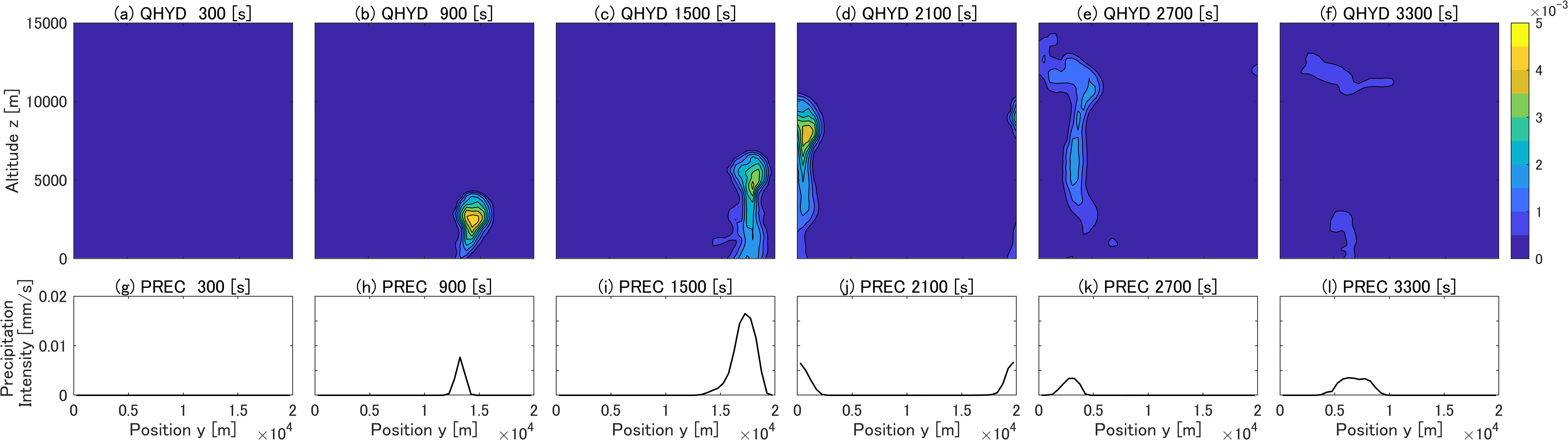}
    \caption{Time evolution of total condensate distribution (top) and surface precipitation distribution (bottom) \cite{ohtsuka}.}
    \label{fig:warm_bubble}
\end{figure*}

For all weather control experiments in this study, a warm bubble experiment is used as a benchmark. In this experiment, a thermal perturbation (warm bubble) is introduced into a specific location within the domain at the initial time to act as a trigger for cumulonimbus convection. 
The atmospheric model is configured on a two-dimensional vertical $y$-$z$ plane with a horizontal resolution of 500 m and 97 vertical layers. A vertically stretched grid system is employed, with a finer layer thickness in the lower atmosphere, as illustrated in Fig.~\ref{fig:grid}. Based on this grid indexing, the horizontal and vertical indices are defined as $i=1,\dots,40$ and $j=1,\dots,97$, respectively, where $(i,j)=(1,1)$ represents the southernmost grid point at the lowest altitude. Accordingly, the vector of the atmospheric state variables over all $N=3880$ grid points is denoted by $\vec{X}_t \in \mathbb{R}^{3880 n_{var}}$, where $n_{var} = 11$ in the present setting, and the vector of the accumulated precipitation over $m = 40$ horizontal grid points is denoted by $\vec{y}_t \in \mathbb{R}^{40}$, respectively. The model top is set at an altitude of 20 km, and periodic boundary conditions are applied to the lateral boundaries.
The initial vertical atmospheric and wind profiles are prescribed based on references \cite{Redel,Ooyama}. Specifically, a warm bubble with a horizontal radius of 4 km and a vertical radius of 3 km is introduced at the center of the domain with a maximum temperature perturbation of 3 K.
The simulation subsequently observes the spatial development of convective clouds and the resulting precipitation distribution over a specified period. 
The total simulation time is 3600 seconds (1 hour). It is divided into $T=12$ intervals of 300 seconds (5 minutes), yielding control instants $t = 0, \dots, T$. Control inputs are applied over $T_c = 10$ control instants, $t=0,\ldots, T_c -1$, because the control input near the end of the weather event does not affect the accumulated precipitation. 
In this paper, a control step refers to the computation and application of one control input at a control instant. 
The computational latency in a control step is ignored in the simulation.  
Figure~\ref{fig:warm_bubble} illustrates the time evolution of total hydrometeors (QHYD) in the atmosphere and the corresponding surface precipitation distribution in six distinct time steps. As depicted in the upper panels of Fig.~\ref{fig:warm_bubble}, the introduction of a warm bubble induces the gradual development of a moisture-rich air mass. Concurrently, as shown in the lower panels, precipitation is triggered and falls in the vicinity of this developing convective system.

\section{Control Methods}\label{sec3}

\subsection{Control Objectives}
This section formulates precipitation control methods via sparse convex optimization of control inputs, using numerically constructed sensitivity matrices. 
The control objectives are the following. 
\begin{enumerate}
    \item Reduce the accumulated precipitation at the terminal time of simulation below a given upper bound whenever possible. If the specified reduction is infeasible, minimize the violation of the constraint. 
    \item Minimize the $\ell_1$ norm of the control input, which is implemented as a perturbation to the atmospheric state, to suppress its magnitude and spatial spread. 
\end{enumerate}
The first objective is essential to mitigate heavy rainfall disasters. However, in general, it is difficult to guarantee a specified level of precipitation reduction in large-scale and complex weather events. Therefore, the optimization problem should be formulated to generate the best possible solution even when the specified constraint is infeasible. 
The second objective to obtain a sparse control input is also important because intervention over a large domain of the atmosphere is difficult. 

To achieve the control objectives, four control methods are formulated: 
(i) initial-only open-loop optimal control (IO-OL), (ii) full-horizon open-loop optimal control (FH-OL), (iii) shrinking-horizon optimal control (SHOC) with fixed terminal time, and 
(iv) single-move MPC with a fixed prediction-horizon length. 
Although variants of the first three methods have been applied to precipitation control in previous studies, they are reformulated in this paper so that their specifications are aligned with the above control objectives. 
Moreover, a tailored MPC formulation is proposed to achieve the above control objectives with a much shorter prediction horizon than the duration of the entire event to reduce the computational cost in the NWP model.

\subsection{Limiting the Perturbation Domain}
To solve optimization problems in weather control, it is essential to quantify the sensitivity---specifically, the impact of perturbations applied to atmospheric states on the resulting accumulated precipitation. 
However, calculating the sensitivity for every grid point in a high-resolution NWP model is computationally prohibitive. 
To enhance computational efficiency, this study adopts the following domain reduction strategies.

First, the vertical range for the perturbation is restricted on the basis of characteristics of meteorological phenomena. 
Although the model top in the warm bubble experiment is set to 20 km, clouds primarily form within the troposphere, reaching altitudes of approximately 15 km at most during summer. 
Furthermore, the vertical extent of cumulonimbus clouds, which are the primary drivers of extreme rainfall, is typically around 10 km even at their mature stage. 
Therefore, the target layers for perturbations are restricted to the bottom 60 layers, covering an altitude range from 0 to approximately 9 km.
Following this reduction, the effective grid dimensions are $i=1,\dots,40$ horizontally and $j=1, \dots, 60$ vertically. 

\begin{figure}[htbp]
    \centering
    \includegraphics[width=0.5\linewidth]{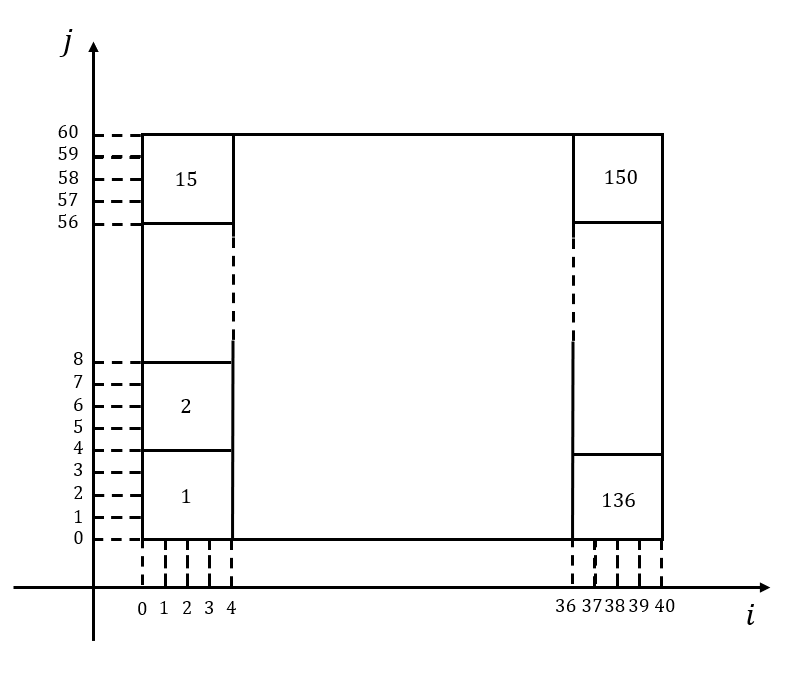}
    \caption{Schematic representation of spatial grid aggregation for sensitivity calculation.}
    \label{fig:grid_aggregation}
\end{figure}

Moreover, this study groups $4 \times 4$ grid cells into a single aggregated block, as illustrated in Fig.~\ref{fig:grid_aggregation}. 
The sensitivity is then computed by perturbing each block as a single unit. 
In this configuration, the total number of control blocks is reduced to $n=150$, which corresponds to less than 4\% of the original $N=3880$ grid points.
This aggregation reduces the computational cost compared to conventional approaches \cite{ohtsuka, KIM} that applied perturbations to each of the $40 \times 97$ grid cells individually, which requires massive computation. 
In the practical implementation of precipitation control, the size of the block needs to be chosen through a trade-off between the computational cost and the realizability of effective intervention in the block, which is beyond the scope of this conceptual study. 

\subsection{Sensitivity Analysis}
Since the NWP model is too complex to use directly for optimization, an approximate linear model around a nominal case is constructed via numerical sensitivity analysis and then used to optimize perturbations of the atmospheric state as the control input. 
The atmospheric state in the nominal case with no perturbation is denoted by $\vec{X}_t^{nom}$ $(t=0,1,\ldots,T)$, and the corresponding accumulated precipitation is denoted by $\vec{y}_{t:t_2}^{nom}$ $(0 \le t < t_2 \le T)$, respectively. 
These nominal values are used in the formulations of the control methods. 

Based on the blocking strategy described in the previous section, let 
$\vec{u}_t$ denote the block-wise control input vector, whose $k$-th component represents a uniform perturbation to $\rho \theta$ in block $k$ at time $t$. 
Let $B \in \mathbb{R}^{n_{var} N \times n}$ be a matrix that maps the block-wise control input $\vec{u}_t$ to the corresponding perturbation of the $\rho \theta$ components in the full vector of atmospheric state $\vec{X}_t^{nom}$. The remaining state components are not directly perturbed. 
Thus, the perturbed state is represented as $\vec{X}_t^{nom} + B \vec{u}_t$. 
If the input $\vec{u}_t$ is sufficiently small, the corresponding output can be approximated by a linear model: 
\begin{equation}
    \vec{y}_{t:t_2} = F_{t:t_2}^{sim}(\vec{X}_t^{nom} + B \vec{u}_t) \approx \vec{y}_{t:t_2}^{nom} + \mathbf{S}_{t:t_2}(\vec{X}_t^{nom}) \vec{u}_t, \label{eq:LM}
\end{equation}
where 
\begin{equation}
    \mathbf{S}_{t:t_2}(\vec{X}_t^{nom}) = \dfrac{\partial F_{t:t_2}^{sim}}{\partial \vec{X}}(\vec{X}_t^{nom})B
\end{equation}
is the sensitivity matrix (Jacobian matrix) of the accumulated precipitation with respect to the control input. 

For finite-difference approximation of the sensitivity matrix, let $\vec{e}_k$ denote a unit vector with only the $k$-th component $1$ and other components $0$, and let $\Delta \vec{y}_{t:t_2}^k$ be the perturbation of the accumulated precipitation corresponding to a finite-difference perturbation vector $d \vec{e}^k$, where $d>0$ is a small scalar: 
\begin{align}
    \Delta \vec{y}_{t:t_2}^k &= F_{t:t_2}^{sim}(\vec{X}_t^{nom} + d B \vec{e}_{k}) - F_{t:t_2}^{sim}(\vec{X}^{nom}_{t}),
\end{align}
for $k=1, \dots, n$. 
Then the sensitivity matrix is approximated by 
\begin{equation}
    \mathbf{S}_{t:t_2} (\vec{X}_t^{nom}) \approx \frac{1}{d}
    \begin{bmatrix}
        \Delta \vec{y}_{t:t_2}^1 & \Delta \vec{y}_{t:t_2}^2 & \cdots & \Delta \vec{y}_{t:t_2}^n
    \end{bmatrix}. \label{eq:FDS}
\end{equation}
The linear model (\ref{eq:LM}) is used to predict the output with the sensitivity matrix approximated by (\ref{eq:FDS}).

Based on the calculated sensitivity matrices, the optimal control input required to achieve the target precipitation level is computed. 
In this study, to ensure spatial sparsity of the perturbation in the atmospheric state, we solve a convex optimization problem that penalizes the $\ell_1$ norm of the control input. The mathematical formulations for each control strategy are described in the following.

\subsection{Initial-Only Open-Loop Optimal Control}
As a baseline, we formulate an IO-OL problem, in which the perturbation is applied only at the initial time $t=0$. 
The convex optimization problem is formulated as follows:
\begin{align}
    \min_{\vec{u}_0, \vec{s}} \quad  \|\vec{u}_0\|_1 + (\alpha \vec{y}_{0:T}^{nom})^\top\vec{s}  
    \quad \text{subject to} \quad  \vec{y}_{0:T}^{nom} + \mathbf{S}_{0:T}(\vec{X}_0^{nom}) \vec{u}_0 \le \vec{y}^{ub} + \vec{s}, \vec{s} \ge 0, \label{eq:IO-OL}
\end{align}
where the inequality constraint is imposed component-wise. 
The first term in the objective function promotes the spatial sparsity of the control input $\vec{u}_0$, i.e., the perturbation in the initial atmospheric state, and the second term penalizes the slack variable $\vec{s}$ representing the constraint violation. 
The constraint imposes an upper bound on the accumulated precipitation at the terminal time. 
In the objective function, the slack variable $\vec{s}$ is weighted by the product of a positive scalar $\alpha$ and the accumulated precipitation $\vec{y}_{0:T}^{nom}$ in the nominal case. 
The accumulated precipitation is nonnegative and serves as a weight to suppress the constraint violation strongly when the nominal accumulated precipitation is large. 
This is a reasonable strategy for mitigating a heavy-rainfall disaster. 

The upper bound $\vec{y}^{ub} \in \mathbb{R}^{m}$ for the accumulated precipitation is defined by applying specific reduction rates to the nominal value. 
Let $y^{max}$ denote the maximum value in the components of $\vec{y}_{0:T}^{nom}$. 
Then, the $i$-th component of $\vec{y}^{ub}$ is set as:
\begin{equation}
    \vec{y}_i^{ub} = \max \{ r \vec{y}_{0:T,i}^{nom}, \, l y^{max} \}, \label{eq:y_ub}
\end{equation}
where the positive parameter $r < 1$ denotes the desired reduction rate for accumulated precipitation, and the positive parameter $l < 1$ acts as a relaxation coefficient, preventing the constraints from becoming overly strict in naturally low-precipitation regions. 

To solve the convex optimization problem (\ref{eq:IO-OL}), the sensitivity matrix $\mathbf{S}_{0:T}(\vec{X}_0^{nom}) \in \mathbb{R}^{m \times n}$ needs to be constructed numerically by finite-difference approximation, which requires $n$ runs of the NWP model with perturbed initial conditions over the entire horizon $\{0, \ldots, T\}$ in addition to the nominal case. 
Once the sensitivity matrix is constructed, a globally optimal solution of the convex optimization problem can be obtained efficiently, which is an advantage of using the linear model. 
However, the control performance depends on the validity of the linear model over the entire horizon. 
In particular, a large reduction in precipitation using only the initial control input may require a large control input; therefore, the linear approximation may no longer be accurate.

\subsection{Full-Horizon Open-Loop Optimal Control}
One possible approach to improve the control performance of the IO-OL method is to leverage the degrees of freedom in the sequence of control inputs over multiple time steps, thereby formulating a FH-OL problem. 
The length $T_c$ for the sequence of control inputs does not need to be identical to the terminal time $T$ of the simulation because precipitation may not be controlled effectively near the end of the weather event. 
The convex optimization problem for FH-OL is formulated as follows:
\begin{align}
    \min_{\vec{u}_0, \cdots, \vec{u}_{T_c-1}, \vec{s}} \quad  \sum_{t=0}^{T_c-1} \|\vec{u}_t\|_1 + (\alpha \vec{y}_{0:T}^{nom})^\top\vec{s}  
    \quad \text{subject to} \quad  \vec{y}_{0:T}^{nom} + \sum_{t=0}^{T_c-1} \mathbf{S}_{t:T}(X_t^{nom}) \vec{u}_t \le \vec{y}^{ub} + \vec{s}, \vec{s} \ge 0, \label{eq:FH-OL}
\end{align}
where the accumulated precipitation at the terminal time is predicted by the linear combination of perturbations caused by the control input (state perturbation) $\vec{u}_t$ and is bounded by the given upper bound $\vec{y}^{ub}$. 
The first term in the objective function promotes the spatial and temporal sparsity of the sequence of state perturbations $\vec{u}_t$ $(t=0,\ldots,T_c-1)$, and the second term penalizes the constraint violation according to the magnitude of the nominal accumulated precipitation. 
A feasible solution to the IO-OL problem in (\ref{eq:IO-OL}) obviously gives a feasible solution to (\ref{eq:FH-OL}) with $\vec{u}_t = 0$ for $t=1,\ldots,T_c-1$. 
Therefore, the minimum value of the objective function in (\ref{eq:FH-OL}) does not exceed that of the objective function in (\ref{eq:IO-OL}). 

This increase in degrees of freedom for control comes at the cost of higher computational demand of the FH-OL problem (\ref{eq:FH-OL}) than the IO-OL problem in (\ref{eq:IO-OL}). 
To solve the convex optimization problem (\ref{eq:FH-OL}), the sequence of the sensitivity matrices $\mathbf{S}_{t:T}(X_t^{nom}) \in \mathbb{R}^{m \times n}$ $(t=0,\ldots,T_c-1)$ needs to be constructed numerically by finite-difference approximation, which requires $n T_c$ runs of the NWP model with perturbed initial conditions over a set of different horizons $\{0, \ldots, T\}, \ldots, \{T_c-1,\ldots, T\}$ in addition to the nominal case. 
Moreover, the control performance still depends on the validity of the linear model over the entire horizon, and the model mismatch cannot be compensated based on the actual response of the NWP model.

\subsection{Shrinking-Horizon Optimal Control}
To further improve the control performance of the FH-OL method, the control input $\vec{u}_t$ at each control instant $t$ should be re-optimized based on the actual atmospheric state $\vec{X}_t$. 
This re-optimization is formulated as an optimal control problem over a shrinking horizon from $t$ to the fixed terminal time $T$. 
Since the re-optimization of the entire sequence $\vec{u}_{t} \ldots,\vec{u}_{T_c-1}$ of the control inputs is computationally demanding, only a single control input $\vec{u}_t$ is optimized in the present formulation. 
This formulation can be viewed as a closed-loop version of the IO-OL method. 
The convex optimization problem at time $t$ in SHOC is defined as follows: 
\begin{align}
    \min_{\vec{u}_t, \vec{s}} \quad \|\vec{u}_t \|_1 + (\alpha \vec{y}_{0:T}^{nom})^\top\vec{s} 
    \quad \text{subject to} \quad  F_{t:T}^{sim}(\vec{X}_t) + \mathbf{S}_{t:T}(\vec{X}_t) \vec{u}_t \le \vec{y}_{t}^{ub} + \vec{s}, \vec{s} \ge 0, \label{eq:SH-OC}
\end{align}
where $\vec{X}_t$ denotes the actual atmospheric state affected by the past control inputs $\vec{u}_0, \ldots, \vec{u}_{t-1}$. 
The inequality constraint specifies the upper bound on the accumulated precipitation predicted from the current state $\vec{X}_t$ over the shrinking horizon. 
Therefore, the present formulation results in closed-loop control utilizing online observation of the atmospheric state. 
The objective function consists of the $\ell_1$ norm of the perturbation $\vec{u}_t$ and the penalty on the slack variable representing the constraint violation. 
The slack variable is weighted by the nominal accumulated precipitation, as in the previous open-loop control methods, IO-OL and FH-OL. 

If the upper bound $\vec{y}_{t}^{ub}$ is set to the same constant value $\vec{y}^{ub}$ as the IO-OL problem, the SHOC problem at $t=0$ is equivalent to the IO-OL problem. 
In this case, the constraint on the upper bound is satisfied only by the initial input $\vec{u}_0$, and small control inputs after the initial time are used only to compensate deviations from the linear model. 
To suppress aggressive interventions during the early stages, the upper bound $\vec{y}_{t}^{ub}$ is defined as a time-varying version of (\ref{eq:y_ub}), guided by a time-varying reduction rate $r_t$ that tightens as time increases:
\begin{align}
    \vec{y}_{t,i}^{ub} = \max \{ r_t \vec{y}_{0:T,i}^{nom}, \, l y^{max} \},  r_t = r + (1 - r)\left(1 - \frac{t+1}{T_c}\right)^3, \label{eq:y_t_ub}
\end{align}
where a positive parameter $r < 1$ denotes the terminal value of the reduction rate at the end of control at $t=T_c - 1$. 
The above reduction rate $r_t$ smoothly decreases from a value close to $1$ at $t=0$ to $r$ at $t = T_c - 1$. 

At each control instant $t$ in SHOC, the current atmospheric state $\vec{X}_t$ is observed, and the updated nominal value of the accumulated precipitation $F_{t:T}^{sim}(\vec{X}_t)$ and the sensitivity matrix $\mathbf{S}_{t:T}(\vec{X}_t)$ are computed, which requires $n+1$ runs of the NWP model from $t$ to the fixed terminal time $T$. 
Then, the convex optimization problem (\ref{eq:SH-OC}) can be solved efficiently. 
The present formulation also generates the sequence of control inputs of length $T_c$ over the simulation, as in the FH-OL method. 
A notable difference is that the sensitivity matrix $\mathbf{S}_{t:T}(\vec{X}_t)$ over the shrinking horizon is constructed only once at each control instant $t$ while the FH-OL method computes all sensitivity matrices over the entire horizon at the initial time. 
Therefore, the computational cost of SHOC is distributed over the control steps and decreases as the time increases due to the shrinking horizon.

\subsection{Model Predictive Control}
To further reduce the computational cost of SHOC while preserving the advantages of closed-loop control, an MPC framework is proposed. 
In particular, a single-move MPC method with a short prediction horizon is formulated to reduce the computational cost of sensitivity analysis, which is dominant in the present precipitation control problem. 
Since the control objectives involve reducing the accumulated precipitation at the terminal time, MPC needs to be carefully formulated to reduce the terminal accumulated precipitation by the repeated optimization over a much shorter horizon. 

At each control instant $t$ in MPC, the control input $\vec{u}_t$ is optimized based on the current state $\vec{X}_t$ by solving the following convex optimization problem: 
\begin{align}
    \min_{\vec{u}_t,\vec{s}} \quad  \|\vec{u}_t \|_1 + (\alpha \vec{y}_{0:T}^{nom})^{\top} \vec{s}  
    \quad \text{subject to} \quad  F_{t:t+L}^{sim}(\vec{X}_t) + \mathbf{S}_{t:t+L}(\vec{X}_t) \vec{u}_t \le \vec{y}_{t:t+L}^{ub} + \vec{s}, \vec{s} \ge 0,\label{eq:MPC}
\end{align}
where $L$ denotes the prediction-horizon length in MPC, and $\vec{X}_t$ denotes the actual atmospheric state affected by the past control inputs $\vec{u}_0, \ldots, \vec{u}_{t-1}$ of MPC. 
The inequality constraint specifies the upper bound on the accumulated precipitation predicted from the current state $\vec{X}_t$ over the prediction horizon from $t$ to $t+L$. 
Therefore, the present formulation also results in closed-loop control, as in SHOC. 
The objective function has the same form as that of SHOC; the main difference lies in the prediction horizon. 

To reduce the terminal accumulated precipitation to a level comparable to that specified by (\ref{eq:y_ub}), the upper bound $\vec{y}_{t:t+L}^{ub}$ for MPC is defined as follows: 
\begin{equation}
    \vec{y}_{t:t+L,i}^{ub} = \max \{ r \vec{y}_{t:t+L,i}^{nom}, \, l y_{t:t+L}^{max} \}, \label{eq:y_ub_MPC}
\end{equation}
where $y_{t:t+L}^{max}$ denotes the maximum value in the components of $\vec{y}_{t:t+L}^{nom}$. 
The present MPC formulation does not directly impose the constraint on the accumulated precipitation at the terminal time. Instead, it imposes a local constraint on the accumulated precipitation over a moving prediction horizon as a computationally tractable surrogate. 
This local upper bound is an empirical surrogate designed to encourage terminal precipitation reduction while keeping the prediction horizon short. 
The actual accumulated precipitation needs to be evaluated at the terminal time in the NWP simulation. 

At each control instant $t$ in MPC, the current atmospheric state $\vec{X}_t$ is observed, and the updated nominal value of the accumulated precipitation $F_{t:t+L}^{sim}(\vec{X}_t)$ and the sensitivity matrix $\mathbf{S}_{t:t+L}(\vec{X}_t)$ are computed over a prediction horizon of length $L$. 
The update of the nominal value and sensitivity matrix requires $n+1$ runs of the NWP model from $t$ to $t+L$, which is much shorter than the entire horizon if $L$ is small. 
Then, the control input $\vec{u}_t$ is determined by solving the convex optimization problem (\ref{eq:MPC}) and applied to the current atmospheric state. 
The present formulation also generates the sequence of perturbations of length $T_c$ over the simulation, as in the FH-OL and SHOC methods. 
A notable difference is that the sensitivity matrix $\mathbf{S}_{t:t+L}(\vec{X}_t)$ over the fixed prediction-horizon length is constructed only once at each control instant $t$. 
Therefore, the uniform computational cost, which depends mainly on $L$, is distributed over the control steps regardless of the entire horizon for simulation. 
The prediction-horizon length $L$ is a crucial design parameter for MPC. 
If $L$ is too small, MPC may not be able to effectively reduce the accumulated precipitation over the prediction horizon due to the delay in the dynamics of the weather event. 
The impact of $L$ on the control performance will be investigated in the next section.

\section{Numerical Experiments}\label{sec4}
\subsection{Computational Environment}
The computational times reported in this section were measured on a computer equipped with an Intel(R) Core(TM) i7-1360P CPU at 2.2 GHz and 32 GB of RAM, running Ubuntu 22.04.5 LTS.
The NWP model, SCALE-RM, is written in Fortran 95 and was compiled using the gfortran compiler.
The numerical experiments for the four control methods were implemented in Python 3.10, and the convex optimization problems were solved using CVXPY. 
No parallel computation was used for the sensitivity analysis.

\subsection{Comparison of Control Methods}
\begin{figure*}[t]
    \centering
    \includegraphics[width=0.8\linewidth]{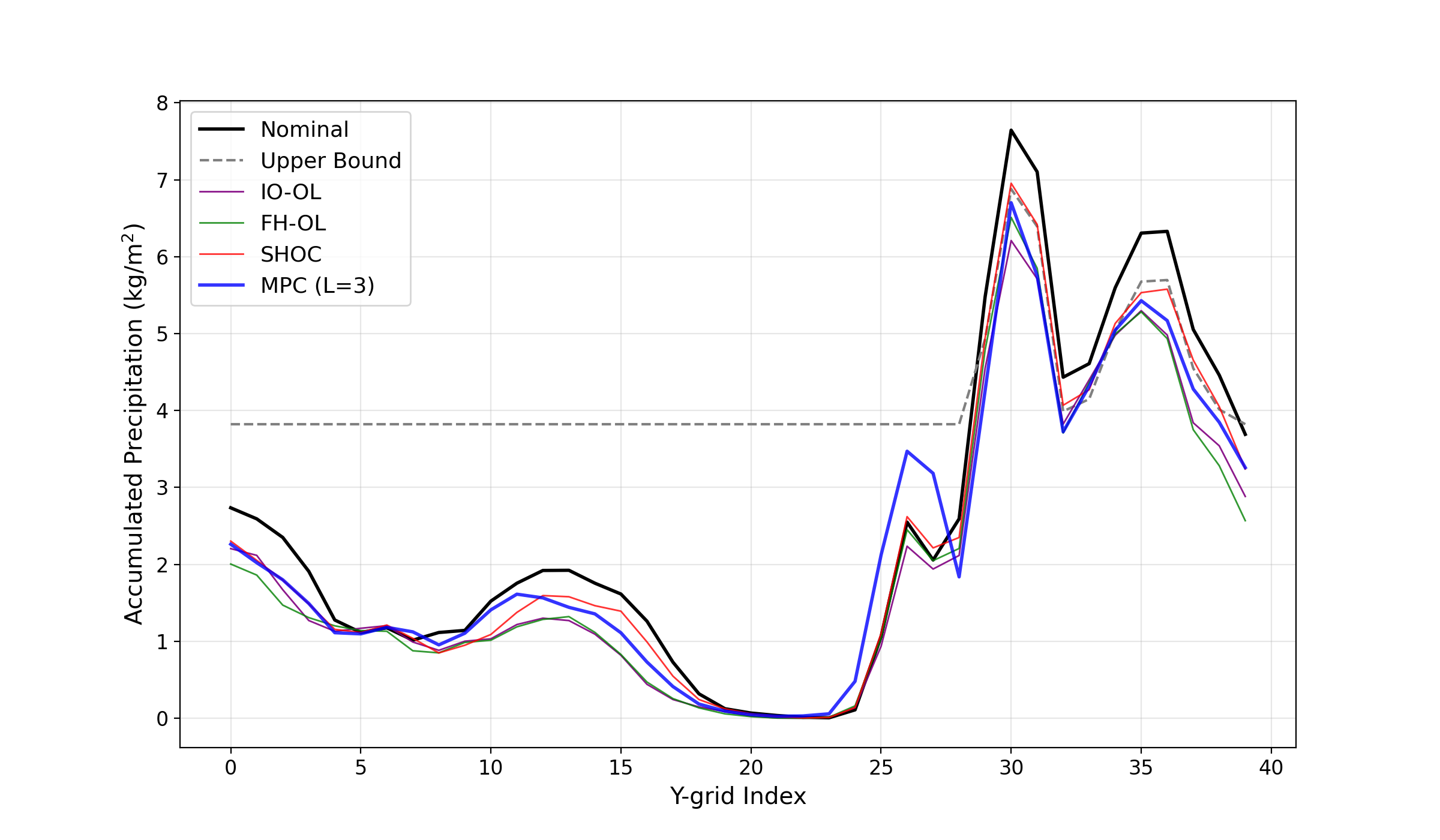}
    \caption{Accumulated precipitation at the terminal time $t=12$ for the reduction rate $r=0.9$.}
    \label{fig:precip_90}
\end{figure*}

\begin{figure*}[t]
    \centering
    \includegraphics[width=0.8\linewidth]{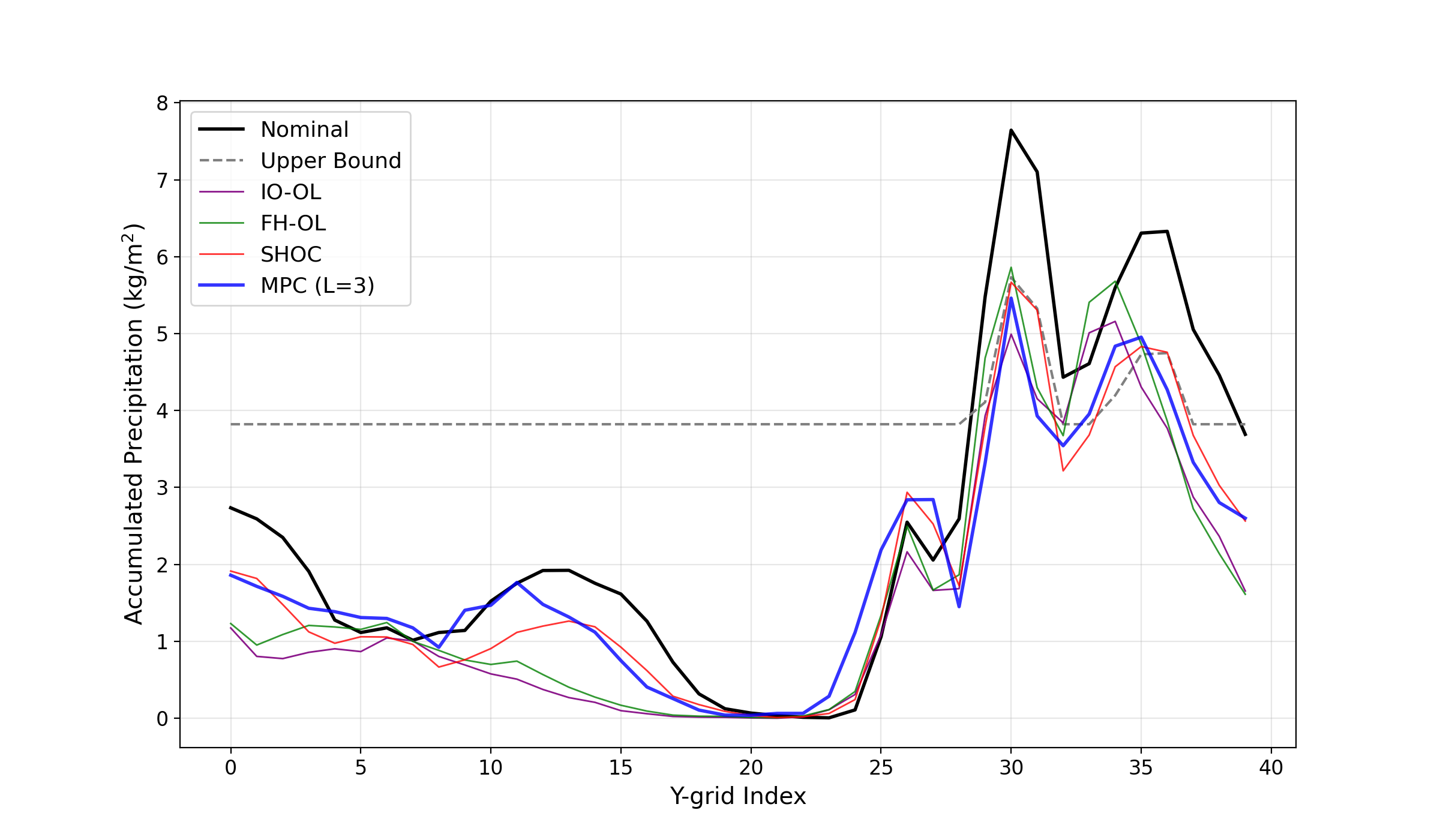}
    \caption{Accumulated precipitation at the terminal time $t=12$ for the reduction rate $r=0.75$.}
    \label{fig:precip_75}
\end{figure*}

\begin{figure}[htbp]
    \centering
    \includegraphics[width=0.4\linewidth]{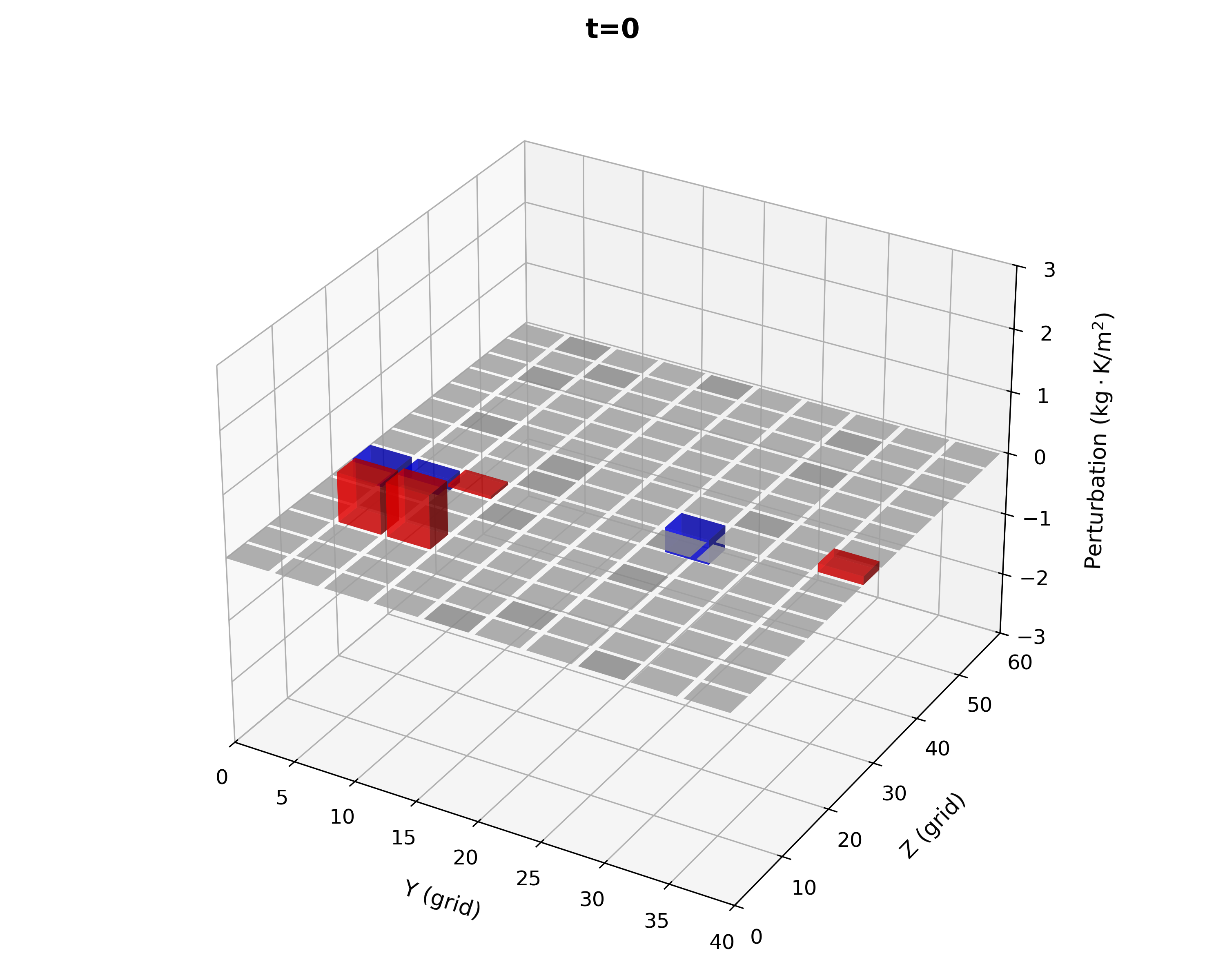}
    \caption{Perturbation to the density-weighted potential temperature in IO-OL for the reduction rate $r=0.9$.}
    \label{fig:pert_init_90}
\end{figure}

\begin{figure*}[htbp]
    \centering
    \includegraphics[width=\linewidth]{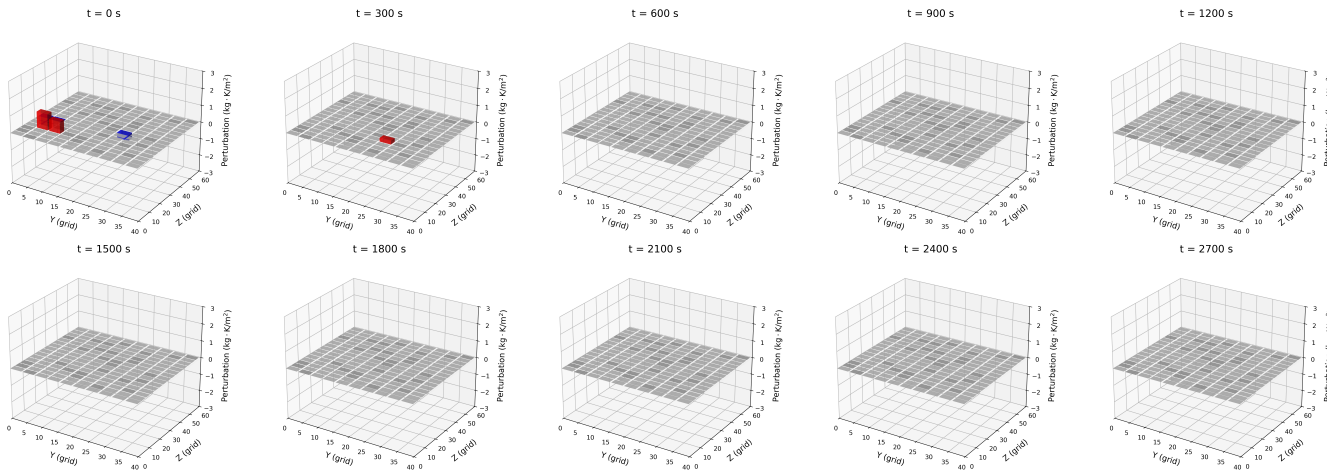}
    \caption{Perturbations to the density-weighted potential temperature in FH-OL for the reduction rate $r=0.9$.}
    \label{fig:pert_90_ff}
\end{figure*}

\begin{figure*}[htbp]
    \centering
    \includegraphics[width=\linewidth]{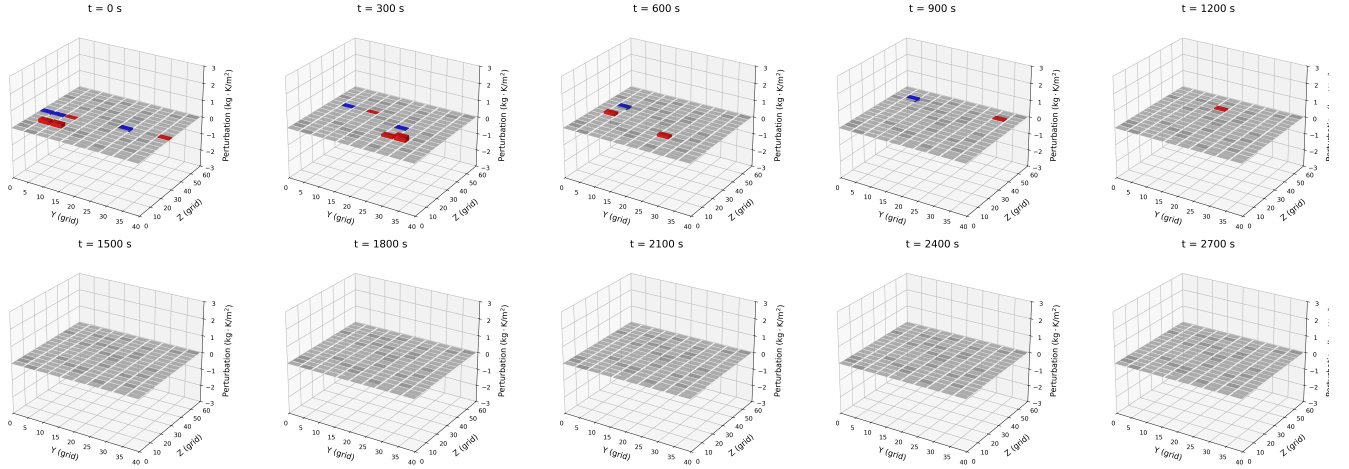}
    \caption{Perturbations to the density-weighted potential temperature in SHOC for the reduction rate $r=0.9$.}
    \label{fig:pert_90_fb}
\end{figure*}

\begin{figure*}[htbp]
    \centering
    \includegraphics[width=\linewidth]{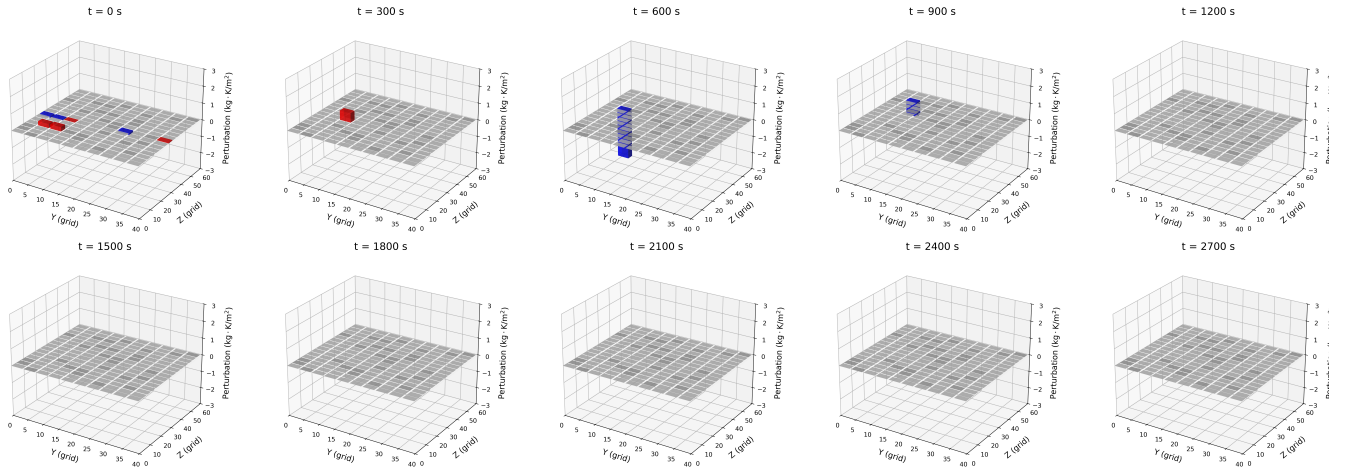}
    \caption{Perturbations to the density-weighted potential temperature in MPC with $L=3$ for the reduction rate $r=0.9$.}
    \label{fig:pert_90_mpc}
\end{figure*}

\begin{figure}[htbp]
    \centering
    \includegraphics[width=0.4\linewidth]{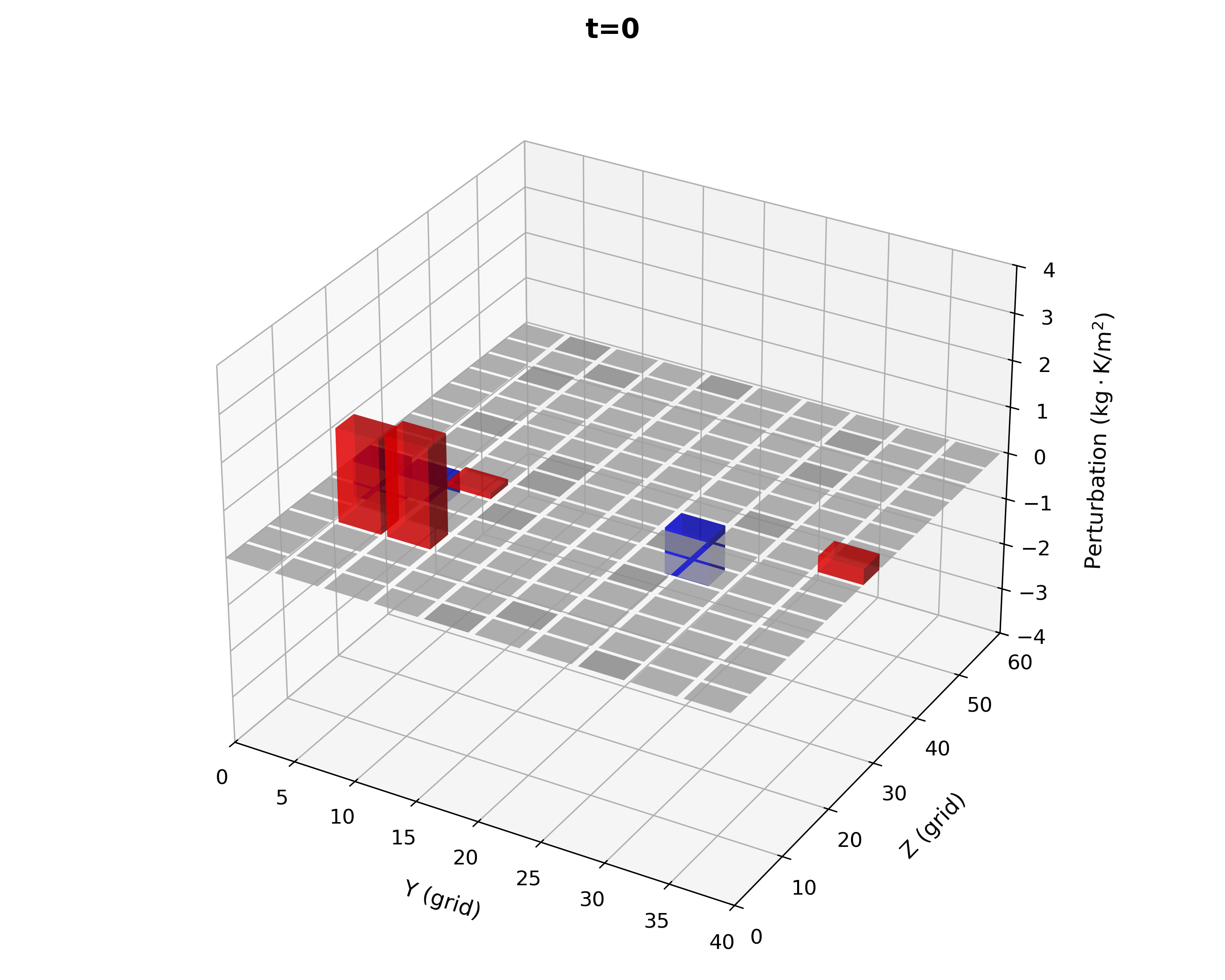}
    \caption{Perturbation to the density-weighted potential temperature in IO-OL for the reduction rate $r=0.75$.}
    \label{fig:pert_init_75}
\end{figure}

\begin{figure*}[htbp]
    \centering
    \includegraphics[width=\linewidth]{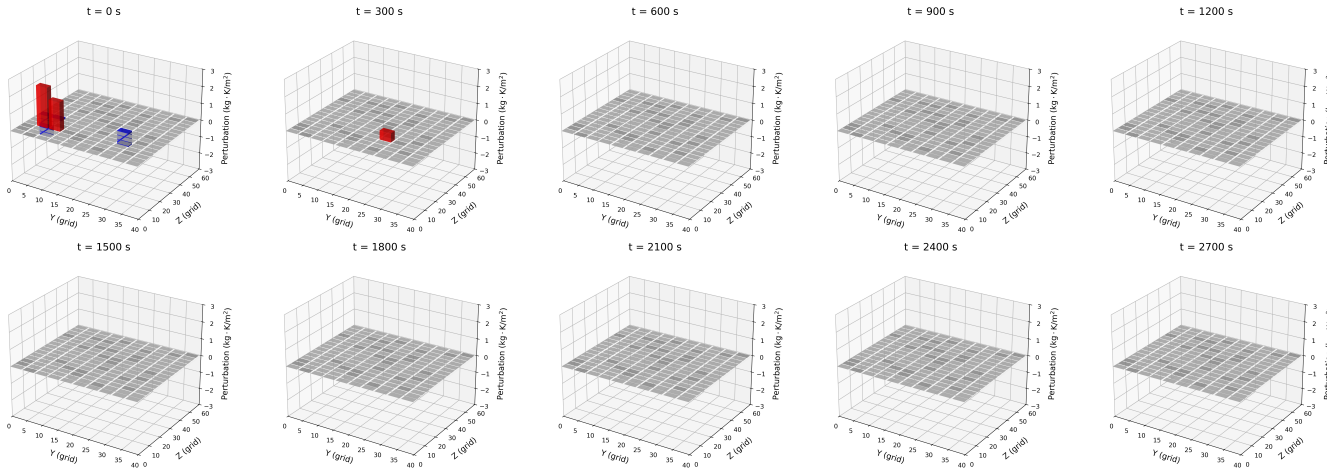}
    \caption{Perturbations to the density-weighted potential temperature in FH-OL for the reduction rate $r=0.75$.}
    \label{fig:pert_75_ff}
\end{figure*}

\begin{figure*}[htbp]
    \centering
    \includegraphics[width=\linewidth]{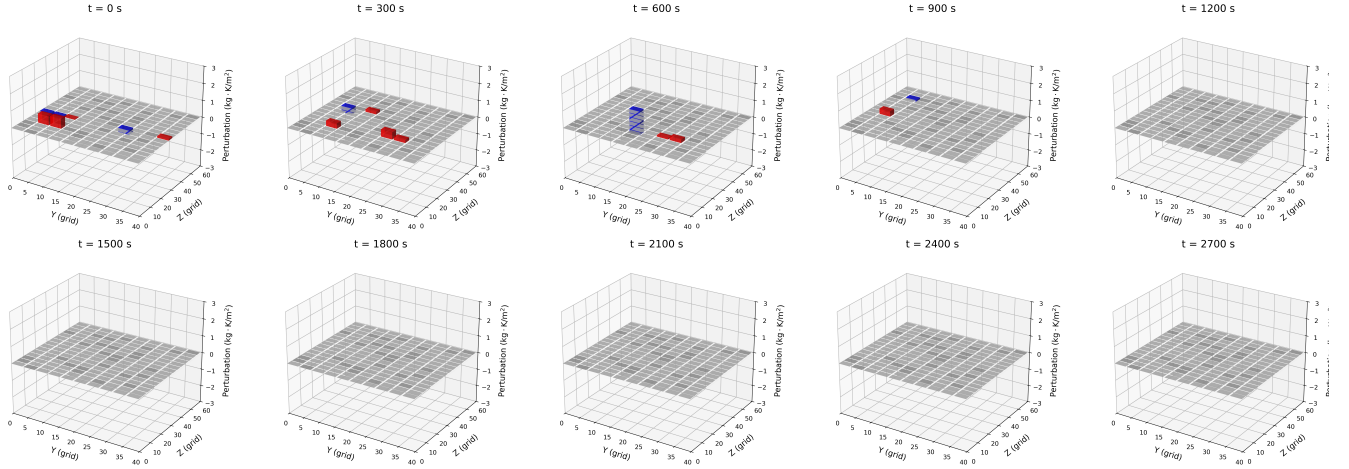}
    \caption{Perturbations to the density-weighted potential temperature in SHOC for the reduction rate $r=0.75$.}
    \label{fig:pert_75_fb}
\end{figure*}

\begin{figure*}[htbp]
    \centering
    \includegraphics[width=\linewidth]{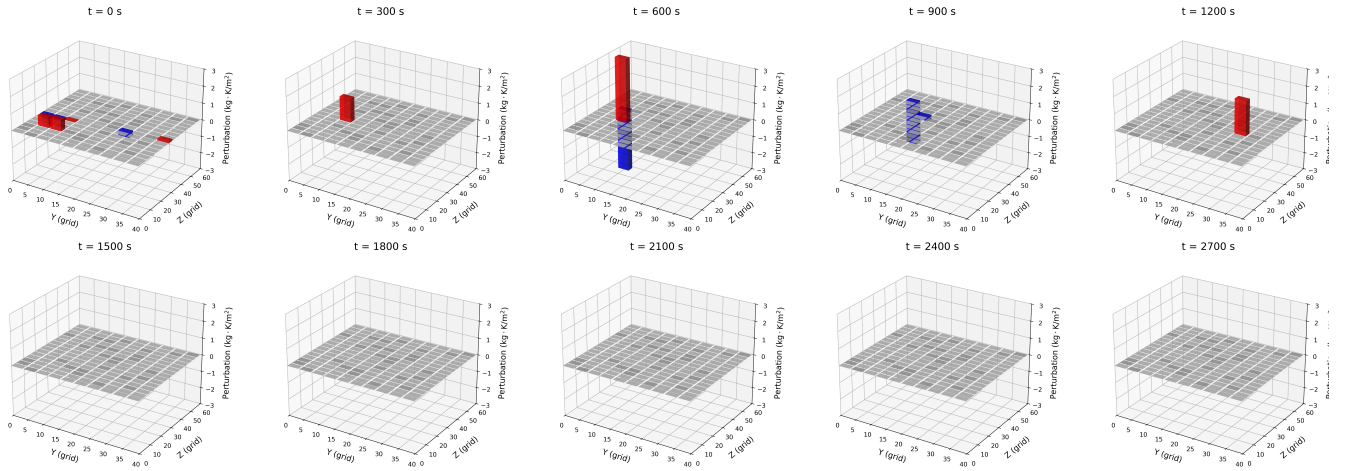}
    \caption{Perturbations to the density-weighted potential temperature in MPC with $L=3$ for the reduction rate $r=0.75$.}
    \label{fig:pert_75_mpc}
\end{figure*}

In this section, we present the results of applying the four control methods: (i) IO-OL method, (ii) FH-OL method, (iii) SHOC method with a fixed terminal time, and (iv) single-move MPC with a fixed prediction-horizon length. 
Two different scenarios are investigated, with the reduction rate set at $r=0.9$ and $r=0.75$, corresponding to 90\% and 75\% of the nominal accumulated precipitation, respectively.  
The relaxation coefficient was set at $l=0.5$ for the upper bounds of the accumulated precipitation in (\ref{eq:y_ub}), (\ref{eq:y_t_ub}), and (\ref{eq:y_ub_MPC}). 
The weight of the slack variable was set at $\alpha=0.15$ in all cases. 
The step size for the finite-difference approximation of the sensitivity matrix was chosen as $d=0.02$ kg$\cdot$K/m$^3$. 
Specifically, we compare and evaluate the control performance based on the achieved spatial distribution of precipitation, the magnitude of control input required at each control instant, and the overall computational cost. 
The corresponding results are summarized in Figs.~\ref{fig:precip_90}--\ref{fig:pert_75_mpc} and Table~\ref{tab:comp_time_detailed}.

\begin{table*}[t]
    \centering
    \caption{Computational time for sensitivity calculations and optimization at each control step. Total times were calculated from unrounded measurements.}
    \label{tab:comp_time_detailed}
    \begin{small} 
    \begin{tabular}{lccccccccccc}
        \hline
        \textbf{Control Method} & \multicolumn{10}{c}{\textbf{Execution Time per Step [min]}} & \textbf{Total Time} \\ 
        & $t=0$ & $t=1$ & $t=2$ & $t=3$ & $t=4$ & $t=5$ & $t=6$ & $t=7$ & $t=8$ & $t=9$ & \textbf{[min]} \\ \hline
        IO-OL & 24.6   & ---   & ---   & ---   & ---   & ---   & ---   & ---   & ---   & ---   & 24.6 \\
        FH-OL & 174.0   & ---   & ---   & ---   & ---   & ---   & ---   & ---   & ---   & ---   & 174.0 \\
        SHOC  & 23.1   & 22.2   & 20.9   & 19.8   & 15.5   & 14.5   & 13.8   & 13.1   & 12.6   & 11.9   & 167.3 \\
        MPC ($L=3$)  & 13.4   & 14.7   & 12.5   & 12.4   & 11.6   & 11.5   & 12.9   & 12.7   & 13.2   & 11.5   & 126.5 \\ \hline
    \end{tabular}
    \end{small}
\end{table*}

First, we discuss the changes in the spatial distribution of precipitation achieved by each method.
In the relatively modest control scenario with a reduction rate $r=0.9$ in Fig.~\ref{fig:precip_90}, all four methods generally satisfied the specified upper bound (dashed line), with no significant performance differences observed among them. 
In contrast, for the reduction rate $r=0.75$ in Fig.~\ref{fig:precip_75}, which requires a more substantial reduction in precipitation, clear differences emerged due to the inherent characteristics of each control strategy.

Although the IO-OL and FH-OL methods successfully suppressed the peak of the accumulated precipitation, an adverse side effect is observed in Fig.~\ref{fig:precip_75}; precipitation inadvertently increased at surrounding grid points, particularly around grid points 33 and 34. 
Since these open-loop control methods rely entirely on optimization performed only at the initial time, they are unable to compensate for cumulative errors caused by the highly nonlinear temporal evolution of the atmospheric system. 
This limitation suggests that such open-loop approaches yield undesirable outcomes when aggressive precipitation control is required.

Conversely, the SHOC method acquires the latest atmospheric state at each control instant and sequentially recomputes the optimal control inputs by re-linearizing the precipitation response around that state. 
This closed-loop compensation allowed the SHOC method to achieve successful reduction of the accumulated precipitation while effectively preventing unintended rainfall increases in the surrounding region around the precipitation peak, as shown in Fig.~\ref{fig:precip_75}. 
Similarly, MPC, which re-optimizes the control input over a short prediction horizon, demonstrated a satisfactory level of precipitation reduction comparable to that of the SHOC method. 

Next, we examine the magnitude and spatial distribution of the control inputs across the different methods. 
In the IO-OL and FH-OL methods in Figs.~\ref{fig:pert_init_90}, \ref{fig:pert_90_ff}, \ref{fig:pert_init_75}, and \ref{fig:pert_75_ff}, which optimize the control inputs based only on the initial state, the control inputs tended to be concentrated in specific regions. 
This concentration likely occurs because these open-loop control methods attempt to account for future response relying solely on initial information, leading to an excessive localized control burden as they try to secure the entire necessary control effort in a single plan.

In contrast, SHOC and MPC in Figs.~\ref{fig:pert_90_fb}, \ref{fig:pert_90_mpc}, \ref{fig:pert_75_fb}, and \ref{fig:pert_75_mpc} demonstrated that the perturbation magnitude required per control step is kept smaller compared to the IO-OL and FH-OL methods. 
By compensating online for model mismatch through re-optimization based on the latest atmospheric state, these closed-loop control methods effectively distribute the control inputs both temporally and spatially. 

Finally, we compare the computational costs across the different control methods: the time required for sensitivity calculations and optimization. 
Table~\ref{tab:comp_time_detailed} presents the breakdown of the execution time in each control step along with the total computational time.
The IO-OL method, which performs optimization exclusively at the initial time ($t=0$), requires only a single optimization of the initial control input. 
This results in the shortest total computational time of 24.6 minutes. However, as previously discussed, its inability to compensate for nonlinear state evolution renders it unsuitable for aggressive precipitation reduction scenarios. 
On the other hand, the FH-OL method, which determines the perturbations for all control instants simultaneously at the initial time, requires solving a massive optimization problem that spans the entire control input sequence. 
Consequently, it requires a substantial computational time of 174.0 minutes for a single execution.

In contrast, SHOC and MPC, which employ sequential re-optimization, exhibit distinct temporal trends in their computational costs. 
The total computational time for the SHOC method is large, reaching 167.3 minutes. 
However, examining the execution time per step reveals a monotonic decrease from 23.1 minutes at $t=0$ to 11.9 minutes at $t=9$. 
This downward trend occurs because the SHOC method utilizes a fixed-terminal-time formulation; as the steps progress, the remaining time to the terminal step shortens, which subsequently reduces both the computational cost for sensitivity analysis and the overall scale of the optimization problem. 

Meanwhile, MPC achieves a total computational time of 126.5 minutes, representing a reduction of 27\% and 24\% relative to FH-OL and SHOC, respectively. 
Furthermore, its execution time per control step remains relatively uniform, taking 12.6 minutes per step on average. 
This reduction in computational cost is mainly attributed to the fixed short prediction horizon used in MPC. 
Consequently, the computational cost of computing the sensitivity matrices at each control instant remains relatively uniform.

These results indicate that the proposed MPC substantially reduces the computational burden relative to SHOC while maintaining comparable precipitation control performance. Although the present serial implementation does not yet satisfy the real-time latency requirement of 5 minutes in the warm bubble experiment, the bounded per-step computational cost suggests a promising direction toward online implementation when combined with parallel sensitivity calculations.

\subsection{Impact of Prediction Horizon Length}
\begin{figure*}[t]
    \centering
    \includegraphics[width=0.8\linewidth]{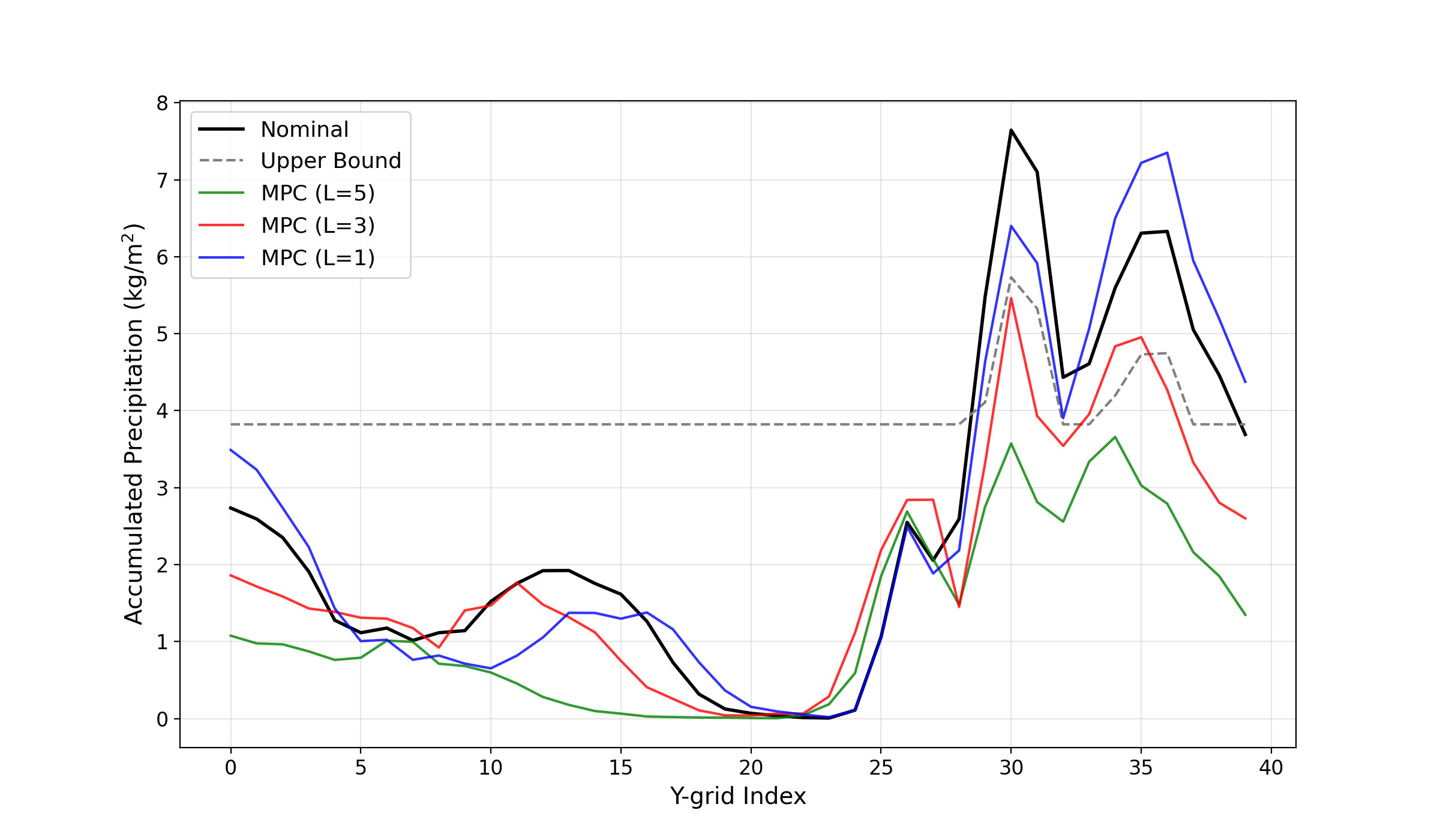}
    \caption{Accumulated precipitation at the terminal time $t=12$ by MPC with different prediction horizons $(L=1, 3, 5)$ for the reduction rate $r=0.75$.}
    \label{fig:mpc_horizon}
\end{figure*}

To further investigate the impact of the prediction horizon in MPC, additional experiments with different horizon lengths were conducted. Here, we compare the baseline MPC ($L=3$) with two additional settings, namely $L=1$ and $L=5$. 
The average computational time per control step was 10.0 minutes for $L=1$ and 13.1 minutes for $L=5$, respectively, while it was 12.6 minutes for $L=3$. 
Figure~\ref{fig:mpc_horizon} shows the spatial distributions of accumulated precipitation obtained using these three prediction horizons. 

The results indicate that the control performance is strongly affected by the choice of prediction-horizon length. When the horizon was shortened to $L=1$, MPC did not sufficiently reduce the accumulated precipitation. 
This is likely because the prediction horizon was too short for the applied control inputs to produce a sufficiently large effect on precipitation. As a result, the optimization problem could not adequately evaluate the future influence of the control input, leading to insufficient precipitation reduction.

In contrast, when the prediction horizon was extended to $L=5$, MPC reduced the accumulated precipitation well below the specified upper bound. 
This result suggests that a longer prediction horizon can improve control performance, although it slightly increases the computational cost in the present experiment. 
Since MPC with $L=3$ also satisfies the upper bound with only a small violation, the prediction-horizon length of $L=3$ is long enough to capture the delayed influence of the control input on precipitation. 
Further increasing $L$ can improve reduction performance but increases the computational cost. 

\section{Conclusions}\label{sec5}

In this study, we developed quantitative precipitation control methods that integrate a realistic NWP model with sensitivity analysis and sparse convex optimization. 
Specifically, we formulated four optimal control problems---IO-OL, FH-OL, SHOC, and MPC---using finite-difference sensitivity matrices and an $\ell_1$-norm penalty to promote spatial sparsity of the control inputs. 
Through the warm bubble benchmark, we compared and evaluated the control performance and computational cost of the four control methods.

The numerical experiments provided important insights into the handling of atmospheric nonlinear dynamics. 
In a demanding control scenario where the target upper bound was set to 75\% of the nominal accumulated precipitation, the open-loop control methods, namely IO-OL and FH-OL, failed to compensate for cumulative errors caused by the highly nonlinear temporal evolution of the atmospheric state. 
In contrast, the closed-loop control methods, namely, SHOC and MPC, iteratively re-optimize the control input based on the atmospheric state, successfully compensating for model mismatch. 
The results demonstrate that the proposed MPC framework can reduce the computational burden of closed-loop precipitation control while maintaining performance comparable to SHOC in the present warm-bubble benchmark. The method should be regarded as a proof-of-concept for NWP-based receding-horizon precipitation control. 

To further validate the effectiveness and applicability of the proposed framework, future work will involve numerical studies using more complex, large-scale three-dimensional simulation environments that model real-world extreme weather events, such as typhoons and LPZs. 
Furthermore, to move toward real-time implementation, a further reduction in computational latency is essential. 
Possible approaches include parallelizing sensitivity analysis and developing a spatiotemporal filtering technique to narrow down the permissible perturbation regions. 
Overcoming these computational and engineering challenges will mark a significant step toward the ultimate goal of proactive disaster mitigation through weather control engineering.

\bmsection*{Conflict of interest}
The authors declare no conflict of interest.

\bmsection*{Data Availability Statement}
Data that support the findings of this study are available from the corresponding author, Toshiyuki Ohtsuka, upon reasonable request.

\bibliography{wileyNJD-AMA}

\end{document}